\documentclass{article}

\usepackage{PRIMEarxiv}

\usepackage[utf8]{inputenc} 
\usepackage[T1]{fontenc}    
\usepackage{hyperref}       
\usepackage{url}            
\usepackage{booktabs}       
\usepackage{amsfonts}       
\usepackage{nicefrac}       
\usepackage{microtype}      
\usepackage{lipsum}
\usepackage{fancyhdr}       
\usepackage{graphicx}       
\graphicspath{{media/}}     
\usepackage{ragged2e}
\usepackage{paralist}
\usepackage{hyperref}
\hypersetup{
colorlinks = true, 
linkcolor=black, 
citecolor=black, 
urlcolor=black 
}

\title{Software Testing in the Quantum World
\thanks{All authors contributed equally. } 
}

\author{
  Rui Abreu \\
  Faculty of Engineering of University of Porto, Portugal; and INESC-ID\\
  Porto, Portugal \\
  \texttt{rui@computer.org} \\
   \And
  Shaukat Ali \\
  Simula Research Laboratory\\
  Oslo, Norway \\
  \texttt{shaukat@simula.no} \\
   \And
  Paolo Arcaini \\
  National Institute of Informatics \\
  Tokyo, Japan \\
  \texttt{arcaini@nii.ac.jp} \\
  \And
 Jos\'{e} Campos \\
 Faculty of Engineering of University of Porto, Portugal; and \\
 LASIGE, Faculdade de Ci\^{e}ncias, Universidade de Lisboa \\
 Lisbon, Portugal \\
  \texttt{jcmc@fe.up.pt} \\
\And
 Michael Felderer \\
German Aerospace Center (DLR), Germany; and University of Cologne \\
Cologne, Germany \\
  \texttt{michael.felderer@dlr.de} \\
\And
Claude Gravel \\
Toronto Metropolitan University \\
Toronto, Canada \\
  \texttt{gravel@torontomu.ca} \\
\And
Fuyuki Ishikawa \\
National Institute of Informatics\\
 Tokyo, Japan  \\
  \texttt{f-ishikawa@nii.ac.jp} \\
\And
Stefan Klikovits \\
Johannes Kepler University\\
 Linz, Austria  \\
  \texttt{stefan.klikovits@jku.at} \\
\And
Andriy Miranskyy \\
Toronto Metropolitan University \\
Toronto, Canada \\
  \texttt{avm@torontomu.ca} \\
  \And
Anila Mjeda \\
Munster Technological University\\
 Cork, Ireland  \\
  \texttt{anila.mjeda@mtu.ie} \\
  \And
Mohammad Reza Mousavi \\
Kings College London\\
 London, UK  \\
  \texttt{mohammad.mousavi@kcl.ac.uk} \\
  \And
Masaomi Yamaguchi \\
Fujitsu Limited, Kawasaki, Japan / The University of Electro-Communications\\
 Tokyo, Japan  \\
  \texttt{y.masaomi@fujitsu.com} \\
  \And
{Lei} Zhang \\
University of Maryland, Baltimore County\\
 United States  \\
  \texttt{leizhang@umbc.edu} \\
  \And
Jianjun Zhao \\
Kyushu University\\
 Fukuoka, Japan  \\
  \texttt{zhao@ait.kyushu-u.ac.jp} \\ 
}

\begin{document}
\maketitle


\centering \textbf{Abstract}

\justifying Quantum computing offers significant speedups for simulating physical, chemical, and biological systems, and for optimization and machine learning. As quantum software grows in complexity, the classical simulation of quantum computers, which has long been essential for quality assurance, becomes infeasible. This shift requires new quality-assurance methods that operate directly on real quantum computers. This paper presents the key challenges in testing large-scale quantum software and offers software engineering perspectives for addressing them.

\section{Introduction}

Miranskyy and Zhang~\cite{TestingQPNIER} highlighted the importance of quantum software testing. Since then, many methods and tools have focused on testing small programs executed on ideal, noise-free classical simulators, conditions rarely met on quantum computers. As quantum computers become powerful enough to simulate classically, larger-scale quantum software must be tested directly on real computers, creating new challenges and research opportunities that this column highlights.

\section{Scale} 

Existing testing methods relying on classical simulation fail to scale owing to exponential state growth, memory constraints, and high computational costs, including hardware-noise simulations. These methods are largely infeasible on real quantum computers, where access is limited, costly, and the hardware is noisy. Many assumptions in the current methods further restrict them to simulators (e.g., requiring many measurements). This mirrors the evolution of classical software testing, where growing complexity has led from exhaustive reasoning to abstract methods (e.g., model-based testing and surrogate models). Therefore, developing test abstractions is essential.

Quantum circuit simplification is a promising approach for improving scalability by reducing circuits and slicing subcircuits as surrogate models for validating software properties. Property-based testing complements this by focusing on properties such as symmetries, invariants, or unitary relations rather than exhaustive output checks. However, the systematic identification of relevant properties remains challenging.

Assume-guarantee decomposition supports scalability by decomposing global properties into component-level contracts, enabling compositional reasoning and targeted integration testing. Together, abstraction, property-based testing, and compositional reasoning form plausible solutions. A key challenge is tool development: fully quantum testing tools do not yet exist, classical tools fail to scale, and hybrid quantum-classical solutions are a near-term goal, whereas fully quantum implementations remain a long-term objective. 

\section{Test Oracle}

Efficient test oracles remain an open challenge~\cite{murillo2025quantum}. Classical strategies based on input-output fail to scale owing to exponential state growth, limited observability, and costly hardware access. Measurement-induced disturbance further complicates oracle design, requiring repeated program runs to obtain statistically meaningful evidence. Shadow tomography allows partial estimation; however, exhaustive inspection remains infeasible,  highlighting the need to shift from deterministic output verification to probabilistic, property-based correctness.

Promising directions emphasize implicit and relational oracles that validate properties rather than explicit outputs~\cite{abreu2022metamorphic}. Property-based oracles check semantic properties such as invariants, symmetries, unitarity, or equivalence between program behaviors, often as relation-checking problems, enabling automated test artifact construction via metamorphic transformations, self-consistency checks, or equivalence testing. Approximate and statistical oracles balance the cost with confidence guarantees, incorporating adaptive sampling strategies and noise-aware thresholds to remain robust under hardware imperfections. Entropy- and distribution-based checks provide additional fault detection without full-state reconstruction, albeit at increased measurement cost. Crucially, these approaches address the quantum kernel in isolation. A scalable oracle strategy must extend across the entire hybrid architecture, shifting the verification from component-level properties to the integrity of the classical-quantum interaction in an end-to-end computational process.
\section{Test Adequacy}

A suitable notion of {\it test adequacy} for quantum programs should shift from \textit{``have the tests exercised all paths?''}, as used in classical computing, to \textit{``have the tests accumulated enough evidence, with stated statistical confidence, that the observed behavior matches the specification within an acceptable noise margin?''}  Given the probabilistic outputs and imperfect hardware, adequacy must capture what was exercised and the confidence with which deviations could be detected.

Building on this, the adequacy of hybrid quantum-classical systems should span the full classical-quantum workflow covering control flow, quantum resources (e.g., qubits, gate types, entanglement structure, and depth), and measurement interfaces that drive classical decisions. In essence, coverage must address both the classical decision surface and the quantum state-preparation/measurement space that it influences.

Given these criteria, the {\it test strength} can be evaluated using fault-based sensitivity (e.g., mutation-style fault injection~\cite{QMutPy}) and statistical power~\cite{miranskyy2025feasibilityquantumunittesting}: the smallest change in the output distribution, expectation value, or performance metric that a test suite can reliably detect at a fixed shot budget. This depends on realistic fault models that combine software defects (e.g., incorrect gates, qubits, basis, or parameters) with execution faults (e.g., decoherence and gate errors).

To achieve adequate efficiency, input-space sampling should be adaptive: begin with diverse seeds, measure coverage and discrepancy, then iteratively select new inputs or parameters to maximize the number of uncovered targets or fault-detection coverage, stopping once coverage goals and confidence stabilize.

\section{Quantum Computing for Quantum Software Testing} 

Software testing involves computationally intensive tasks, such as test generation, prioritization, and selection, over large input spaces. As software grows in size and complexity, classical techniques increasingly rely on heuristics and trading off optimality to achieve scalability. Beyond revisiting testing concepts, a complementary question is whether quantum computing can serve as both a test target and a computational resource for testing itself.

Quantum computing offers a distinct computational model with potential speedups for certain search, optimization, and learning problems that align naturally with core testing activities. Quantum search can accelerate the discovery of test inputs that satisfy coverage or fault conditions, complementing classical constraint-based generation. Test prioritization and minimization problems can be formulated as combinatorial optimization tasks, amenable to hybrid quantum--classical approaches, such as quantum approximate optimization or annealing. Quantum machine learning also enables data-driven testing tasks such as fault localization and approximation of test oracles without precise specifications.

In the near term, these approaches are quantum-assisted rather than fully quantum and are integrated selectively with hybrid workflows alongside classical heuristics. Despite current hardware limitations, this perspective clarifies how, in the long term, quantum computing could provide an additional computational layer to reduce the growing cost of software testing tasks as technology matures~\cite{Miranskyy2022using}.

\section{Benchmarks and Tools} 
Benchmarking is central to assessing the scalability, effectiveness, and practical relevance of quantum software testing techniques. Unlike classical benchmarks, quantum testing benchmarks must account for hardware noise, finite shot budgets, calibration drift, and backend heterogeneity, as these factors directly influence observed behavior. Meaningful benchmarks should therefore report not only fault-detection rates or coverage, but also resource usage (e.g., shots, depth, qubits), statistical confidence, and robustness to noise and calibration variability. This creates a natural connection between software and hardware benchmarks: low-level hardware benchmarks (e.g., gate fidelity, readout error) may be reused to parametrize and contextualize software-level testing results. 

Community coordination around shared benchmark suites and experimental protocols is essential. Public, curated benchmarks with representative programs, standardized fault models, and clearly defined metrics enable fair comparison across tools and platforms, support reproducibility through transparent reporting of execution parameters, and allow longitudinal evaluation as hardware evolves.

Although several tools exist (see Wang et al.~\cite{wang2025landscape}), a key question is whether they can scale and perform effectively on real quantum hardware. Tool fragmentation, hardware constraints, and the probabilistic nature of quantum computers pose significant obstacles.

Tooling remains difficult to integrate into standard IDEs and CI/CD pipelines, owing to limited interoperability, inconsistent APIs, backend variability, and a lack of standardized formats. Solutions include stable IDE plugins, unified test and result specifications, and reference CI/CD templates for managing credentials, noise models, and execution parameters. Interpreting test results is challenging owing to probabilistic and noise-sensitive outputs. Useful reports should include effect sizes, confidence intervals, circuit-diff views, and calibration metadata to aid in diagnosis.

Large language models have the potential to automate test generation and debugging, but they often produce unreliable outputs. Mitigation includes retrieval-grounded copilots, schema-constrained prompting, and benchmarks assessing correctness and reproducibility. Benchmarks must reflect the real hardware and integrate low-level hardware tests with oracles to distinguish software faults from hardware imperfections. Domain-specific benchmarks are also crucial, using representative programs, diverse fault types, and mutation operators tailored to domain-specific patterns and error models.

\section{Conclusions}

The most pressing challenge in quantum software testing is achieving scalable, end-to-end quality assurance on real, noisy quantum computers within a hybrid quantum–classical software setup. Addressing this requires a shift toward abstraction, property-based, and statistically grounded testing. Successfully meeting this challenge will enable the community to produce reliable solutions that achieve quantum advantage while also ensuring their dependability.

\bibliographystyle{unsrt}
\bibliography{references}

@article{wang2025landscape,
  title={The Landscape of Quantum Software Testing Tools},
  author={Wang, Xinyi and Ali, Shaukat and Taibi, Davide},
  journal={IEEE Software},
  volume={42},
  number={5},
  pages={136--140},
  year={2025},
  publisher={IEEE}
}

@inproceedings{TestingQPNIER,
author = {Miranskyy, Andriy and Zhang, Lei},
title = {On testing quantum programs},
year = {2019},
publisher = {IEEE Press},
url = {https://doi.org/10.1109/ICSE-NIER.2019.00023},
doi = {10.1109/ICSE-NIER.2019.00023},
booktitle = {Proceedings of the 41st International Conference on Software Engineering: New Ideas and Emerging Results},
pages = {57–60},
numpages = {4},
location = {Montreal, Quebec, Canada},
series = {ICSE-NIER '19}
}

@article{murillo2025quantum,
  title={Quantum software engineering: Roadmap and challenges ahead},
  author={Murillo, Juan Manuel and Garcia-Alonso, Jose and Moguel, Enrique and Barzen, Johanna and Leymann, Frank and Ali, Shaukat and Yue, Tao and Arcaini, Paolo and P{\'e}rez-Castillo, Ricardo and Garc{\'\i}a-Rodr{\'\i}guez de Guzm{\'a}n, Ignacio and others},
  journal={ACM Transactions on Software Engineering and Methodology},
  volume={34},
  number={5},
  pages={1--48},
  year={2025},
  publisher={ACM New York, NY}
}

@article{QMutPy,
  author={Fortunato, Daniel and Campos, Jos\'{e} and Abreu, Rui},
  journal={IEEE Transactions on Quantum Engineering}, 
  title={{Mutation Testing of Quantum Programs: A Case Study With Qiskit}}, 
  year={2022},
  volume={3},
  number={},
  pages={1-17},
  doi={10.1109/TQE.2022.3195061},
}

@misc{miranskyy2025feasibilityquantumunittesting,
  title={{On the Feasibility of Quantum Unit Testing}},
  author={Andriy Miranskyy and José Campos and Anila Mjeda and Lei Zhang and Ignacio García Rodríguez de Guzmán},
  year={2025},
  eprint={2507.17235},
  archivePrefix={arXiv},
  primaryClass={cs.SE},
  url={https://arxiv.org/abs/2507.17235},
}

@inproceedings{abreu2022metamorphic,
  title={Metamorphic testing of oracle quantum programs},
  author={Abreu, Rui and Fernandes, Jo{\~a}o Paulo and Llana, Luis and Tavares, Guilherme},
  booktitle={Proceedings of the 3rd International Workshop on Quantum Software Engineering},
  pages={16--23},
  year={2022}
}

@inproceedings{miranskyy2022using,
  title={Using quantum computers to speed up dynamic testing of software},
  author={Miranskyy, Andriy},
  booktitle={Proceedings of the 1st International Workshop on Quantum Programming for Software Engineering},
  pages={26--31},
  year={2022}
}


\end{document}